\documentclass[journal]{IEEEtran}
\usepackage{cite}
\usepackage{capt-of}

\ifCLASSINFOpdf
\else
\fi
  \usepackage[caption=false,font=footnotesize]{subfig}

\usepackage{etex}

\usepackage{amsfonts}

\usepackage{stmaryrd}

\usepackage{amsmath}
\usepackage[makeroom]{cancel}
\usepackage{amsthm}

\usepackage[linesnumbered,ruled]{algorithm2e}
\makeatletter
\newcommand{\removelatexerror}{\let\@latex@error\@gobble}
\makeatother
\SetKwRepeat{Do}{do}{while}%

\SetCommentSty{mycommfont}
\usepackage{amssymb}

\usepackage{textcomp}

\usepackage{bm}
\usepackage{pdfpages}
\usepackage[T1]{fontenc}

\usepackage{multirow}
\usepackage{comment}
\usepackage{tikz}
\usetikzlibrary{automata,positioning,shapes,arrows}

\usepackage{array}
\newcolumntype{P}[1]{>{\centering\arraybackslash}p{#1}}

\usepackage[english]{babel}
\addto\captionsenglish{\renewcommand*{\tablename}{TABLE}}

\newtheoremstyle{exampstyle}
{\topsep} 
{\topsep} 
{} 
{} 
{\bfseries} 
{.} 
{.5em} 
{} 
\theoremstyle{exampstyle} 
\theoremstyle{exampstyle} 
\theoremstyle{exampstyle} \newtheorem{proposition}{Proposition}
\theoremstyle{exampstyle} \newtheorem{definition}{Definition}
\theoremstyle{exampstyle} 
\theoremstyle{exampstyle} \newtheorem{problem}{Problem}

{\tiny }

\begin{document}
%
\title{Supervisor Synthesis of POMDP based on Automata Learning}
%
%
%

\author{Xiaobin~Zhang, 
        Bo~Wu, 
        and~Hai~Lin 
\thanks{The authors are with Department of Electrical Engineering, University of Notre Dame, Notre Dame, IN 46556, USA.
	{\tt\small (xzhang11@nd.edu;~bwu3@nd.edu;~hlin1@nd.edu)}.}
\thanks{The financial supports from NSF-EECS-1253488 and NSF-CNS-1446288 for this work are greatly acknowledged.}
}

%
%

\markboth{~}%
{Shell \MakeLowercase{\textit{et al.}}: Bare Demo of IEEEtran.cls for IEEE Journals}
%



\maketitle

\begin{abstract}
As a general and thus popular model for autonomous systems, partially observable Markov decision process (POMDP) can capture uncertainties from different sources like sensing noises, actuation errors, and uncertain environments. However, its comprehensiveness makes the planning and control in POMDP difficult. Traditional POMDP planning problems target to find the optimal policy to maximize the expectation of accumulated rewards. But for safety critical applications, guarantees of system performance described by formal specifications are desired, which motivates us to consider formal methods to synthesize supervisor for POMDP. With system specifications given by Probabilistic Computation Tree Logic (PCTL), we propose a supervisory control framework with a type of deterministic finite automata (DFA), za-DFA, as the controller form. While the existing work mainly relies on optimization techniques to learn fixed-size finite state controllers (FSCs), we develop an $L^*$ learning based algorithm to determine both space and transitions of za-DFA. Membership queries and different oracles for conjectures are defined. The learning algorithm is sound and complete. An example is given in detailed steps to illustrate the supervisor synthesis algorithm.
\end{abstract}	


%
\IEEEpeerreviewmaketitle

\section{Introduction}
%
%
%
%
\IEEEPARstart{I}{n} real world applications, autonomous systems always contain uncertainties. The planning and control problem for such systems has become a hot research area in recent years with background varying from navigation \cite{olson2013exploration, grocholsky2006cooperative}, communication protocol design \cite{zhao2007decentralized}, autonomous driving \cite{sadigh2014data,lam2014pomdp}, and human-robot collaboration \cite{gopalanmodeling,schmidt2008human,zhang2016performance}. Different system models have been considered to capture uncertainties, and partially observable Markov decision process (POMDP) has emerged as one of the most general and thus popular models. POMDP models system software and hardware statuses with discrete states. Between different states, probabilistic transitions are triggered by different system actions to describe uncertainties from the system actuation behavior. Compared to Markov decision process (MDP), POMDP considers partial observability on its states that can model sensing noises and observation errors, which makes MDP a special case of POMDP. This property is very useful in modeling autonomous systems with hidden states, such as advanced driver assistant system (ADAS) \cite{lam2014pomdp} and human-robot collaboration \cite{schmidt2008human,taha2008pomdp} where human intention can not be directly observed. Together with probabilistic transitions between states and nondeterminism in action selections, POMDP can capture uncertainties from various sources, such as sensing, actuation, and the environment. Meanwhile, a reward function can be defined that assigns real value to each state transition to represent additional information in POMDP.

In this paper, we study formal design methods for POMDPs with control tasks given by Probabilistic Computation Tree Logic (PCTL).  While most of the $\omega-$regular properties are undecidable for POMDPs \cite{chatterjee2016decidable}, we consider PCTL specifications with finite horizons which can bound the searching space with finite memory in POMDP model checking following the philosophy of Bounded Model Checking \cite{biere2003bounded}. Meanwhile, a lot of robotics applications require task completion with finite time, such as motion planning \cite{yoo2013provably}, which also makes PCTL specification with finite horizons suitable to describe our control tasks. With a finite planning horizon, the model checking problem of POMDP is decidable, but a history-dependent controller instead of a memoryless one is necessary. To regulate POMDP to satisfy a finite horizon PCTL, we propose a novel supervisor control framework with a special type of deterministic finite automaton (DFA), za-DFA, as the supervisor to achieve history-dependent planning. After defining the probability space of POMDP, PCTL satisfaction over POMDP is established on the product model between za-DFA and POMDP. To check the satisfaction relation efficiently, we show the connection between the model checking and the optimal policy computation, then modify a state-of-art POMDP solving algorithm, Partially Observable Monte-Carlo Planning (POMCP) \cite{silver2010monte}, to reduce the computational complexity for POMDP model checking. After that, an $L^*$ learning based supervisor synthesis algorithm is proposed to synthesize a za-DFA to satisfy the given specification. To guarantee the soundness and completeness of the supervisor synthesis, we design novel algorithms to answer membership queries and conjectures from $L^*$ learning. The returned za-DFA will also be permissive by enabling more than one action for POMDP to select given a history.  

\subsection{Related Work}
Traditional planning and control problems in POMDP target to find a \emph{policy} that maximizes the expectation of accumulated rewards. Since states are not directly observable, the available information for a control policy is an observation-action sequence up to current time instance, and such a sequence is called \emph{history}. History can be represented in a compact form called \emph{belief state}, which is a probability distribution over the state space of POMDP. Since belief state is sufficient statistics for history \cite{astrom1965optimal}, POMDP can be viewed as MDP with a continuous state space formed by belief states. This inspires solving POMDP planning by finding the optimal control policy over the continuous belief state space.  Exact planning of POMDP \cite{sondik1978optimal} can be intractable with the size of state space and planning horizon exploding quickly. Therefore, approximation methods are proposed to approximate the value function or limit the policy search space to alleviate the computational complexity. As one of the most popular approaches, point-based value iteration (PBVI) optimizes the value function only over a selected finite set of belief states and provides the optimization result with a bounded error \cite{cheng1988algorithms, zhang2001speeding,pineau2006anytime, kurniawati2008sarsop}.

Compared to the point-based approach that solves POMDP on the continuous state space of belief states, the controller-based approach \cite{amato2010finite} finds the optimal policy represented by a finite state controller (FSC) with finite memory. An FSC can be defined as a directed graph $G=\langle\mathcal{N},\mathcal{E}\rangle$ with each node $n\in \mathcal{N}$ being labeled by an action $a$ and each edge $e\in\mathcal{E}$ by an observation $z$ in POMDP. Each node has one outward edge per observation, and a policy can be executed by taking action associated with the node at current time instance and updating the current node by following the edge labeled by the observation made\cite{poupart2003bounded}. This representation is equivalent to Moore machine \cite{hill1979introduction} from automata theory\cite{amato2010finite}. There are two types of approaches to find an FSC: policy iteration and gradient search. The policy iteration tends to find the optimal controller, but the size of the controller can grow exponentially fast and turns intractable. The gradient search usually leads to a suboptimal solution that often traps in local optimum \cite{aberdeen2002scaling,meuleau1999solving}. To combine the advantages from gradient ascent and policy iteration, bounded policy iteration (BPI) is proposed in \cite{poupart2003bounded} to limit the size of the controller and provide evidence to help escape local optimum. Besides direct graph as the controller form for FSC, DFA and Mealy machine have also been considered in \cite{grzes2013isomorph} and \cite{amato2010finite}, respectively.

Currently, most existing results on the control problem of POMDP focus on the reward-based planning. However, for some safety-critical applications like autonomous driving, a guaranteed system performance is crucial. This motivates us to consider formal methods. In robotics, formal methods are used to generate controllers that can
guarantee the system performance to satisfy high-level mission requirements \cite{fainekos2005hybrid,kress2009temporal,fainekos2009temporal,wongpiromsarn2013synthesis}. For complicated missions, temporal logic \cite{manna2012temporal} is an efficient tool to describe requirements for system tasks due to its expressiveness and similarity to natural languages. Compared to extensive studies in reward-based planning, very few results on formal methods based planning have been established for POMDP, which makes it an open problem \cite{lahijanian2009probabilistic}. Until recently, there are some advances in the controller synthesis of POMDP under temporal logics. In \cite{sharan2014formal}, the controller synthesis of POMDP with Linear Temporal Logic (LTL) specifications over infinite horizon is discussed and solved based on gradient search where fixed-size FSCs are used to maximize the probability of satisfying the given LTL specification. However, this method suffers from local maxima, and the initial choice of the FSC's structure does not have a systematic guideline \cite{sharan2014formal}. In \cite{chatterjee2015symbolic}, the authors use observation-stationary (memoryless) controller to regulate POMDP to satisfy almost-sure reachability properties. Since the action is selected only depends on current observation, the satisfiability modulo theories (SMT) method is applied with similar idea shows in \cite{junges2016safety} where a state-based controller for MDP is learned. Compared to history-dependent controllers, memoryless controllers used in these work are not general enough for reasoning over finite horizons. In \cite{vasile2016control}, a linear time invariant system with linear observation model for states is considered, which is equivalent to a discrete time continuous space POMDP. The system specification is given as Gaussian Distribution Temporal Logic (GDTL) as an extension of Boolean logic. Sampling-based algorithms are proposed to build a transition system to generate a finite abstraction for belief space. With the specification being converted to deterministic Rabin automaton, the synthesis is done on the product MDP following dynamic programming approach. However, the size of the product MDP still suffers under the curse of history for POMDPs. Similar to POMDPs, the deterministic systems with partial information have also been studied for synthesis problems in \cite{shani2011replanning,shani2013heuristics,fu2016synthesis}. However, applying these methods to POMDP is hard due to the probabilistic transition nature of POMDP. 

Since POMDP is an extended model of MDP, there is also some related work for supervisor synthesis on MDPs. Especially for permissive controller design, in \cite{drager2014permissive}, state-based controllers without memory are proposed for infinite horizon planning and Mixed Integer Linear Programming (MILP) \cite{smith2008tutorial} is applied to find a permissive controller. Similarly, in \cite{junges2016safety}, SMT is combined with reinforcement learning to learn a state-based controller. While these methods assume a memoryless controller, a history-dependent controller is necessary for POMDP planning over a finite horizon. Also due to the partial observability in POMDPs, applying these methods to POMDP are fundamentally difficult. Besides these works, using the $L^*$ algorithm to learn system supervisor has also been considered in our previous work for MDPs \cite{wu2015counterexample}. To apply the $L^*$ algorithm to POMDP supervisor synthesis, in this paper, we extensively discuss the supervisor synthesis framework and design new membership query and conjecture checking rules to overcome the difficulties brought by partial observability.  

\subsection{Our Contributions}
This paper is an extended and revised version of our preliminary conference paper \cite{zhang2015learning}. Compared to \cite{zhang2015learning}, this paper makes the following new contributions. First, we formally build the POMDP supervisor control framework by proving the sufficiency of using za-DFA as the controller form, defining the probability space for POMDP, then establishing PCTL satisfaction over POMDP. Secondly, the model checking of POMDP over observation-based adversaries is intensively studied, and a modified POMCP algorithm is given to conquer the computational complexity. Thirdly, we develop new oracles for the $L^*$ learning algorithm to guarantee the completeness and allow permissiveness of the supervisor. Based on that, a new example is given to illustrate the learning process in detailed steps.  

The technical contributions are summarized in the order in which they appear in the paper as follows:
\begin{itemize}
	\item We propose a supervisory control framework for POMDP to satisfy PCTL specifications over finite horizons. As a special type of DFA, za-DFA is used as the supervisor form. Based on that, we further define the probability space and PCTL satisfaction over POMDP. Then the POMDP model checking is intensively discussed and a modified POMCP method is given to speed up the model checking process.  
	\item We design an $L^*$ learning based supervisor synthesis algorithm to learn a suitable supervisor automatically. With properly defined membership queries and conjectures, our learning algorithm is sound and complete. The returned za-DFA can be permissive, and the non-blocking feature is guaranteed.
\end{itemize}

\subsection{Outline of the Paper}
The rest of this paper is organized as follows. In Section \uppercase\expandafter{\romannumeral2}, MDP-related preliminaries are given with definitions and notations. The supervisory control framework for POMDP is proposed in Section \uppercase\expandafter{\romannumeral3}. Following by that, Section \uppercase\expandafter{\romannumeral4} presents $L^*$ learning based supervisor synthesis algorithm. The analysis and discussions are addressed in Section \uppercase\expandafter{\romannumeral5}. Section \uppercase\expandafter{\romannumeral6} gives an example to illustrate the learning process. Finally, Section \uppercase\expandafter{\romannumeral7} concludes this paper with the future work. 

\section{Preliminaries}
\subsection{MDP Modeling, Paths and Adversaries}

MDPs are probabilistic models for systems with discrete state spaces. With nondeterminisms from decision making and probabilistic behavior in system transitions, MDPs are widely used to model system uncertainties.

\begin{definition}
	\cite{rutten2004mathematical} An MDP is a tuple $\mathcal{M}=(S,\bar{s},A,T)$ where
	\begin{itemize}
		\item $S$ is a finite set of states;
		\item $\bar{s}\in S$ is the initial state;
		\item $A$ is a finite set of actions;
		\item $T:S\times A \times S\rightarrow [0,1]$ is a transition function.
	\end{itemize}
\end{definition}
\noindent Here $T(s,a,s')$ describes the probability of making a transition from a state $s\in S$ to another state $s'\in S$ after taking an action $a\in A$.

In MDPs, there are multiple actions defined for each state. If we limit the number of actions defined for each state to be 1, we have a discrete-time Markov chain (DTMC). 
\begin{definition}
	\cite{rutten2004mathematical} A DTMC is a tuple $\mathcal{M}=(S,\bar{s},T)$ where
	\begin{itemize}
		\item $S$ is a finite set of states;
		\item $\bar{s}\in S$ is the initial state;
		\item $T:S\times S\rightarrow [0,1]$ is a transition function.
	\end{itemize}
\end{definition}

To analyze the behavior of MDP and DTMC with additional information, we can define a labeling function $L: S\rightarrow 2^{AP}$ that assigns each state $s\in S$ with a subset of atomic propositions $AP$. This helps to introduce system requirements in forms of temporal logics.

In MDP $\mathcal{M}=(S,\bar{s},A,T)$, a $path$ $\rho$ is a nonempty sequence of states and actions in the form 
\begin{align*}
	\rho = s_0a_0s_1a_1s_2\ldots
\end{align*}
where $s_0 = \bar{s}, ~s_i\in S,~a_i\in A$ and $T(s_i,a_i,s_{i+1})\geq0$ for all $i\geq0$ \cite{rutten2004mathematical}. Generally, we denote the $i$th state $s_i$ of a path $\rho$ as $\rho(i)$ and the length of $\rho$ (the number of transitions) as $|\rho|$. We use $Path_\mathcal{M}$ to represent the set of all possible paths in $\mathcal{M}$ and $Pref_{\mathcal{M}}$ for its set of corresponding prefixes.

To solve the nondeterminism in MDP, we need an $adversary$ to build a map between system paths and actions. Depending on whether a deterministic action is selected or a probability distribution over all possible actions is given, there are two types of adversaries: pure adversary and randomized adversary. For the pure adversary, it is a function $\sigma:Pref_{\mathcal{M}}\rightarrow A$, that maps every finite path of $\mathcal{M}$ onto an action in $A$. For the randomized adversary, it is a function $\sigma:Pref_{\mathcal{M}}\rightarrow Dist(A)$, which maps every finite path of $\mathcal{M}$ onto a distribution over $A$. With an adversary $\sigma$ that solves the nondeterminism in MDP, the set of possible MDP paths is denoted as $Path^\sigma_\mathcal{M}$ and the regulated system behavior can be represented as a DTMC.

\subsection{PCTL and PCTL Model Checking over MDPs}

For a labeled MDP, we can use PCTL \cite{rutten2004mathematical} to represent the system design requirements. PCTL is the probabilistic extension of the Computation Tree Logic (CTL) \cite{clarke1982design}. 
\begin{definition} \cite{rutten2004mathematical}
	The syntax of PCTL is defined as
	\begin{itemize}
		\item State formula $\phi::=true~|~\alpha~|~\neg \phi~|~\phi\wedge\phi~|P_{\bowtie p}[\psi]$,
		\item Path formula $\psi::=X\phi~|~\phi~\mathcal{U}^{\leq k}\phi~|~\phi~\mathcal{U}~\phi,$
	\end{itemize}
	where $\alpha\in AP$, $\bowtie\in\{\leq,<,\geq,>\}$, $p\in[0,1]$ and $k\in\mathbb{N}$.
\end{definition}
\noindent Here $\neg$ stands for "negation", $\wedge$ for "conjunction", $X$ for "next", $\mathcal{U}^{\leq k}$ for "bounded until" and $\mathcal{U}$ for "until". Specially, $P_{\bowtie p}[\psi]$ takes a path formula $\psi$ as its parameter and describes the probabilistic constraint. 

Given the syntaxes of POMDP, we can define PCTL satisfaction relation on MDP as follows.
\begin{definition}\cite{rutten2004mathematical}
	For an labeled MDP $\mathcal{M}=(S,\bar{s},A,T,L)$, the satisfaction relation $\vDash$ for any states $s\in S$ is defined inductively:
	\begin{align*}
		s &\vDash true,~\forall s\in S;\\
		s & \vDash \alpha \Leftrightarrow \alpha\in L(s);\\
		s &\vDash \neg\phi \Leftrightarrow s\nvDash\phi;\\
		s &\vDash \phi_1\wedge\phi_2 \Leftrightarrow s\vDash\phi_1\wedge s\vDash\phi_2;\\
		s &\vDash P_{\bowtie p}[\psi] \Leftrightarrow Pr(\{\rho\in Path^{\sigma}_\mathcal{M}|~ \rho \vDash\psi \})\bowtie p,~\forall \sigma\in \Sigma_{\mathcal{M}},
	\end{align*}
	where $\Sigma_\mathcal{M}$ is the set of all adversaries and for any path $\rho\in Path_\mathcal{M}$
	\begin{align*}
		\rho &\vDash X\phi \Leftrightarrow \rho(1)\vDash \phi;\\
		\rho &\vDash \phi_1~\mathcal{U}^{\leq k}\phi_2 \Leftrightarrow \exists i\leq k, \rho(i)\vDash \phi_2\wedge\rho(j)\vDash\phi_1,\forall j<i;\\
		\rho &\vDash \phi_1~\mathcal{U}\phi_2 \Leftrightarrow \exists k\geq 0, \rho\vDash\phi_1~\mathcal{U}^{\leq k}\phi_2.
	\end{align*}
\end{definition}

The model checking of PCTL specification has been extensively studied for MDPs \cite{rutten2004mathematical}. PCTL specifications with probabilistic operators are considered. Depending on whether $\bowtie$ in the specification gives lower or upper bound, PCTL model checking of MDPs solves an optimization problem by computing either the minimum or maximum probability over all adversaries \cite{rutten2004mathematical}. Since the states are fully observable, the model checking for MDPs can be solved following dynamic programming techniques with polynomial time complexity \cite{baier2008principles}. Different software tools for MDP model checking are available, such as PRISM \cite{kwiatkowska2011prism} and recently developed model checker Storm \cite{dehnert2017storm}.

\section{POMDP modeling and supervisory control framework}

In this section, we propose a supervisory control framework to regular the close-loop behavior of POMDP to satisfy finite horizon PCTL specifications. 

\subsection{POMDP Modeling, Paths and Adversaries}

POMDPs are widely used to capture systems uncertainties from difference aspects. As an extension of MDP model, POMDP considers states with partial observability to model uncertainties from system sensing. 
\begin{definition}
	A POMDP is a tuple $\mathcal{P}=\{\mathcal{M},Z,O\}$ where
	\begin{itemize}
		\item $\mathcal{M}$ is an MDP;
		\item $Z$ is a finite set of observations;		
		\item $O:S\times Z\rightarrow [0,1]$ is an observation function.
	\end{itemize}
\end{definition}
\noindent In POMDP, the observable information for each state $s\in S$ is given by $O$ as a probability distribution over $Z$. Here $O(s,z)$ stands for the probability of observing $z\in Z$ at state $s\in S$. Then MDP can also be viewed as a special case of POMDP where its $Z=S$ and its observation function defined for each $s\in S$ is a Dirac delta function with 
\begin{align*}
O(s,z)=\begin{cases}
1,~z=s;\\
0,\text{~otherwise}.
\end{cases}
\end{align*}

\noindent\textbf{Remark}: Since states in POMDP are not directly observable, it may happen that we observe an observation $z$ and decide to take action $a$ while $a$ is not defined for the current real state $s$. In this case, no state transitions will be triggered, as the system will ignore this command and stay in its current state. 

Due to the partial observability, paths in POMDP can not be directly observed then used as the information for POMDP planning. Instead, the observation sequence of a path $\rho =s_0a_0s_1a_1s_2\ldots$ can be defined as a unique sequence $obs(\rho)=z_0a_0z_1a_1z_2\ldots$ where $z_i\in Z$ and $O(s_i,z_i)>0$ for all $i\geq0$ (if $obs(\rho_1)\neq obs(\rho_2)$, then $\rho_1$ and $\rho_2$ are considered as different paths). This observation sequence can be seen as history in traditional POMDP planning problems. While history is defined to start with an action, the initial observation $z_0$ in the observation sequence can be seen as a special observation $Init$ for the initial state $\bar{s}$ with $O(\bar{s},Init)=1$ since we assume $\bar{s}$ is known. If the initial status of POMDP is given as a probability distribution over $S$, we can add a dummy initial state then define its transitions to other $s\in S$ based on the initial probability distribution \cite{zhang2015learning}. In the rest of this paper, we will use the observation sequence and history for POMDP interchangeably if the meanings are clear.

Given histories as control inputs, the planning problem of POMDP needs to find an adversary as a mapping function that maps every finite history onto an action in $A$ or a probability distribution over $A$. As in MDP, the former type of adversaries is called pure adversary, and the later is called randomized adversary. As a special case of the randomized adversary, the pure adversary is less powerful generally. But for the finite horizon PCTL specifications considered in our work, the pure adversaries and randomized adversaries have the same power in the sense that restricting the set of adversaries to pure strategies will not change the satisfaction relation of the considered PCTL fragments \cite{chatterjee2010randomness}. While the detailed analysis follows the fact that POMDP is a one-and-a-half player game \cite{chatterjee2010randomness}, the intuitive justification for this claim is that if we are just interested in upper and lower bounds to the probability of some events to happen, any probabilistic combination of these events stays within the bounds. Moreover, pure adversaries are sufficient to observe the bounds \cite{chatterjee2010randomness}. Therefore, we consider the controller design of pure adversary in our supervisory control framework. 

\subsection{Supervisory Control with za-DFA}

We want to find a supervisor to provide pure adversaries for POMDP and regulate the closed-loop behavior to satisfy finite horizon PCTL specifications. To improve the permissiveness, we target to find a set of proper pure adversaries. Since the control objective is given by a finite horizon specification, history-dependent controller outperforms history-independent (memoryless or observation-stationary) one and its justification can be directly inherited from MDP cases \cite{drager2014permissive}. Based on these facts, we propose za-DFA as the supervisor for POMDP with the alphabet being defined in a particular form.

\begin{definition}\cite{zhang2015learning}\label{zaDFA}
	A supervisor for POMDP $\mathcal{P}$$=$$\{S,\bar{s},A,Z,T,O\}$ is a za-DFA $\mathcal{F}$$=$$\{Q,\bar{q},\Sigma,\delta,Q_m\}$, where
	\begin{itemize}
		\item $Q$ is a finite set of states;
		\item $\bar{q}\in Q$ is the initial state;
		\item $\Sigma=\{\alpha=\langle z,a\rangle|~z\in Z,a\in A\}$ is the finite alphabet;
		\item $\delta:Q\times\Sigma\rightarrow Q$ is a transition function;	
		\item $Q_m$ is a finite set of accepting states.
	\end{itemize}
\end{definition}

Since DFA is an equivalent representation of regular language \cite{kumar2012modeling}, za-DFA represents a regular set of strings with the set of the observation-action pairs in POMDP as its alphabet. A path $q_0\langle z_0,a_0\rangle...q_n\langle z_n,a_n\rangle$ in $\mathcal{F}$ is a string concatenation of these pairs, which encodes a history $z_0a_0...z_n$ with an action $a_n$. Then the accepted runs in za-DFA give the enabled actions for different histories and represent POMDP executions. Note that the prefixes of the accepted runs must also be accepted since we have to allow the prefixes to happen in POMDP execution first. This implies that the accepted language $L_m(\mathcal{F})$ of za-DFA $\mathcal{F}$ as the supervisor for POMDP is prefix-closed, i.e., $Pref(L(\mathcal{F}))=L_m(\mathcal{F})$ where $Pref(L(\mathcal{F}))$ denotes all prefixes of the language of $\mathcal{F}$ \cite{kumar2012modeling}. 

\begin{proposition}\label{theorem::zaDFA}
	A set of pure adversaries to regulate a finite horizon PCTL specification for POMDP $\mathcal{P}=\{S,\bar{s},A,Z,T,O\}$ can always be represented as a za-DFA.
\end{proposition}
\begin{IEEEproof}
	A pure adversary in POMDP maps a history $h$ to an action $a\in A$. Since we consider finite POMDP, the observation set $Z$ and action set $A$ are finite, which form a finite alphabet for za-DFA. Meanwhile, for a finite horizon PCTL specification, the pure adversaries give the action selection rules for finite length histories. Thus all possible concatenations of history $h$ and action $a$ enabled by this set of pure adversaries will form a finite set of strings $U$ and each string $y\in U$ has a finite length. Then we can define a nondeterministic finite automaton (NFA) $\mathcal{F}^N$ such that its accepted language is exactly the set $U$. Here $\mathcal{F}^N$ can be constructed by unifying the initial state for DFA representing each string $y\in U$. By applying the subset construction on NFA $\mathcal{F}^N$, we can get a DFA whose accepted language is $U$ \cite{kumar2012modeling}. With the set of observation-action pairs as the alphabet, we have shown that we can always find a za-DFA to represent a set of pure adversaries to regulate a finite horizon PCTL specification for POMDP.    
\end{IEEEproof}

Given a za-DFA as the supervisor for POMDP, all histories that may be encountered during POMDP executions are mapped to a set of enabled actions. Then we can define a product MDP as the parallel composition between POMDP and za-DFA to describe the regulated behavior. 

\begin{definition}\label{def::Mza}
	Given a POMDP $\mathcal{P}=\{S,\bar{s},A,Z,T,O\}$ and a za-DFA $\mathcal{F}=\{Q,\bar{q},\Sigma,\delta,Q_m\}$ as the supervisor, their parallel composition $\mathcal{P}||\mathcal{F}$ is an MDP $\mathcal{M}^\mathcal{F}=(S^\mathcal{F},\bar{s}^\mathcal{F},A^\mathcal{F},T^\mathcal{F})$,
	\begin{itemize}
		\item $S^\mathcal{F}=\{szq|O(s,z)>0,s\in S,z\in Z,q\in Q_m\}\cup\{\bar{s}\bar{q}\}$ is a finite set of states;
		\item $\bar{s}^\mathcal{F}=\bar{s}\bar{q}$ is the initial state;
		\item $A^\mathcal{F}=A$ is a finite set of actions;
		\item $T^\mathcal{F}(\bar{s}\bar{q},a,s'z'q')=O(s',z')T(\bar{s},a,s')$, if $\delta(\bar{q},\langle z,a\rangle)=q'$ with $O(\bar{s},z)>0$, $T(\bar{s},a,s')>0$ and $O(s',z')>0$;
		\item $T^\mathcal{F}(szq,a,s'z'q')$ $=$ $O(s',z')T(s,a,s')$, if $\delta(q,\langle z,a\rangle )=q'$ with $T(s,a,s')>0$ and $O(s',z')>0$.
	\end{itemize}
\end{definition} 
For the labeling function, $L^\mathcal{F}(\bar{s}\bar{q})=L(\bar{s})$ and $L^\mathcal{F}(szq)=L(s),~\forall s\in S,z\in Z,q\in Q_m$.

\noindent\textbf{Remark}: Compared to the global Markov chain defined in \cite{sharan2014formal} describing the regulated behavior of POMDP under an FSC, the product MDP defined in Definition \ref{def::Mza} is more general because za-DFA is permissive and it enables more than one action to be selected under a history. 

To make za-DFA feasible for POMDP planning in practice, we require that a POMDP $\mathcal{P}$ should not get "blocked" under the supervision of $\mathcal{F}$ in the sense that there always exists at least one action being enabled given a history allowed in $\mathcal{F}$.

\begin{definition}\label{def::nonblocking}
	A supervisor za-DFA $\mathcal{F}$ to regulate POMDP $\mathcal{P}$ for a finite horizon $k$ is non-blocking, if there are outgoing transitions defined on all states that are reachable in $k$ steps from $\bar{s}\bar{q}$ in $\mathcal{M}^{\mathcal{F}}=\mathcal{P}||\mathcal{F}$.  	
\end{definition}
\noindent Compared to the feasibility constraint defined in our previous work \cite{zhang2015learning}, here we allow multiple actions being enabled given a history to have permissiveness in the supervisory control framework using za-DFA.

Given a non-blocking za-DFA $\mathcal{F}=\{Q,\bar{q},\Sigma,\delta,Q_m\}$, the simulation run of POMDP $\mathcal{P}=\{S,\bar{s},A,Z,T,O\}$ is shown in Algorithm~\ref{execution}. Starting from initial state $\bar{s}$, $\mathcal{P}$ first generates an observation $z(i)$ on state $s(i)$ at each time instance $i$. Then $\mathcal{F}$ will search for an outgoing transition $\langle z(i),a(i)\rangle$ from $q(i)$ to any $q\in Q_m$ with $q(0)=\bar{q}$ and the corresponding action $a(i)$ is selected to execute. After that, the state of $\mathcal{F}$ is updated and a new POMDP state is simulated following action $a(i)$.  
\begin{figure}[!t]
	\removelatexerror
	\begin{algorithm}[H]
		\DontPrintSemicolon 
		$s(0)\gets \bar{s}$, $q(0)\gets \bar{q}$\;
		\For{$i =0,1,...,k$} {
			simulate $z(i)$ based on $O$ given $s(i)$\;
			choose any $a(i)\in A$ such that $\langle z(i),a(i)\rangle$ defines an outgoing transition from $q(i)$ to any $q\in Q_m$\;
			$q(i+1)\leftarrow \delta(q(i),\langle z(i),a(i)\rangle)$\;
			simulate $s(i+1)$ based on $T$ given $s(i)$ and $a(i)$\;	
		}
		\caption{Simulation run of POMDP $\mathcal{P}$ regulated by za-DFA $\mathcal{F}$ up to time $k$}
		\label{execution}
	\end{algorithm}
		\vspace{-20pt}
\end{figure}

\subsection{Probability Space and PCTL Satisfaction over POMDP}
To formally address the PCTL satisfaction over POMDP, we first define the probability space in POMDP. With an observation-based adversary, the behavior of POMDP is purely probabilistic. Given a finite path $\rho_{fin}$ and its corresponding observation sequence $obs(\rho_{fin})$, with an observation-based adversary, we can define the basic cylinder set in POMDP $\mathcal{P}$ as follows: 
\begin{align*}
	\mathcal{C}(\rho_{fin}):&=\{\rho\in Path_{\mathcal{P}}|~\rho_{fin}\text{ is prefix of }\rho\\
	&~~~~\text{ and }obs(\rho_{fin})(i)=obs(\rho)(i),\forall i \leq |\rho_{fin}|\},
\end{align*}
which is the set of all infinite paths with the prefix $\rho_{fin}$ and observation prefix $obs(\rho_{fin})$. Let $Cyl$ contain all sets $\mathcal{C}(\rho_{fin})$ where $\rho_{fin}$ ranges over all paths with all possible observation sequences. Then the $\sigma$-algebra can be defined on the paths generated by $Cyl$ and the corresponding probability measure can be defined as
\begin{align*}
	&Pr(\mathcal{C}(\rho_{fin}))=\\
	&~~~~\begin{cases}
	1, ~|\rho_{fin}|=0;\\
	O(s(0),z(0))\displaystyle \prod^{|\rho_{fin}|}_{i=1}T(s(i-1),a(i),s(i))O(s(i),z(i)),\\
	~~~~~~~~~~~~~~~~~~~~~~~~~~~~~~~~~~~~~~~~~~~~~\text{otherwise,}
	\end{cases}
\end{align*} 
where $s(i)=\rho_{fin}(i),z(i)=obs(\rho_{fin})(i)$, and $a(i)$ is the selected action from the adversary given the observation sequence up to time instance $i$. Since we assume the initial state $\bar{s}$ is given, the initial observation will be the special observation $Init$ with $O(\bar{s},Init)=1$. With the domain $Path_{\mathcal{P}}$, $\sigma$-algebra and the corresponding probability measure, we have defined the probability space for POMDP under an observation-based adversary. These results are modified based on \cite{zhang2005logic} where the probability space for Hidden Markov Model (HMM) is defined. 

Since PCTL over MDP is well defined, the product MDP that describes the regulated behavior of POMDP under the supervision of za-DFA can be used to connect PCTL satisfaction over POMDP with its definition for MDP. Given a path $\rho$ in $\mathcal{M}^{\mathcal{F}}$, we can have its observation sequence of $obs(\rho)$ by extracting the observation symbol $z$ out of the $szq$ tuple for the state in $\rho$ (the observation symbol for $\bar{s}\bar{q}$ is the special observation $Init$). With an observation-based adversary, there is a one-to-one correspondence between the paths in $\mathcal{M}^{\mathcal{F}}$ and $\mathcal{P}$. Then based on the general definition of the probability space on MDP \cite{rutten2004mathematical}, it is not hard to see that the probability spaces on MDP $\mathcal{M}^{\mathcal{F}}$ and POMDP $\mathcal{P}$ are equivalent. Therefore, given a POMDP $\mathcal{P}$ and a za-DFA $\mathcal{F}$, the PCTL satisfaction with a finite horizon over the regulated system is equivalent to the PCTL satisfaction over the product MDP $\mathcal{M}^{\mathcal{F}}$ constraining to observation-based adversaries. We denote the model checking on $\mathcal{M}^{\mathcal{F}}$ constrained to the observation-based adversaries as $\mathcal{M}^{\mathcal{F}}\models_{obs}\phi$ where $\models_{obs}$ stands for satisfaction relation constrained to the observation-based adversaries.

For the sake of simplicity, we consider bounded until PCTL specification $\phi=P_{\unlhd p}[\phi_1~\mathcal{U}^{\leq k}\phi_2]$ with $\unlhd\in\{\leq,<\}$ in the rest of this paper. But for finite horizon PCTL, the generality is not lost since lots of finite horizon PCTL specifications can be transformed to bounded until form and the model checking mechanism is similar as shown in \cite{han2009counterexample}.

\subsection{POMDP Model Checking}

To verify the satisfaction relation over the regulated behavior, we need to solve the PCTL model checking problem on $\mathcal{M}^{\mathcal{F}}$ where most of the operators are handled in the same way as in MDP model checking. But for state formula $P_{\bowtie p}[\psi]$, we need to check whether the probability bound $\bowtie p$ is satisfied given the observation-based adversaries instead of all adversaries. We can solve this by computing either the minimum or maximum probability depending on whether a lower or upper bound is defined by $\bowtie$ \cite{rutten2004mathematical}. This problem can be solved with EXPTIME-complete complexity for finite horizon specifications. But with the size of POMDP and the planning horizon increasing, this problem becomes much harder to solve. Another promising approach is to convert the model checking to an equivalent optimal policy computation problem on POMDP. Following this method, we can leverage recently developed POMDP solvers that can handle a larger problem size with high-efficiency  \cite{zhang2001speeding,pineau2006anytime,kurniawati2008sarsop}. For the finite horizon PCTL $\phi=P_{\unlhd p}[\phi_1~\mathcal{U}^{\leq k}\phi_2]$, the model checking of this type of specifications can be converted to an optimal policy computation problem by modifying the transition structure of POMDP to make all states $s\models\neg\phi_1$ and states $s\models\phi_2$ absorbing, and designing the reward scheme that assigns 0 to intermediate transitions and 1 to the final transitions on $s\models\phi_2$ when the planning depth $k$ is reached \cite{sharan2014formal,zhang2017counterexample,silver2010monte}.

\begin{figure}[t!]
	\removelatexerror
	\begin{algorithm}[H]
		\DontPrintSemicolon 
		\SetKwFunction{FSimulate}{SIMULATE}
		\SetKwFunction{FTIMEOUT}{TIMEOUT}
		\SetKwFunction{FSearch}{SEARCH}
		\SetKwProg{Fn}{Function}{:}{\KwRet}
		\Fn{\FSearch{$h$}}{
			\Do{not \FTIMEOUT{}}{
				\uIf{$h=empty$}{$s=\bar{s}$}\Else{$s\sim B(h)$}
				\FSimulate{$s,h,0$}\;
			}		
			\KwRet $\underset{a}{\arg\max}~V(ha)$ \;
		}
		\;
		\SetKwFunction{FRoll}{ROLLOUT}
		\SetKwProg{Pn}{Function}{:}{\KwRet}
		\Pn{\FRoll{$s$, $h$, $depth$}}{
			\uIf{$depth>k$}{\KwRet 0}
			\ElseIf{$depth=k$}{\KwRet $\sum_{s\models\phi_2}b(s,h)$}
			$a\sim \sigma_{rollout}(h,\cdot)$\;
			$(s',z)\sim \mathcal{G}(s,a)$\;
			\KwRet \FRoll{$s'$,$haz$,$depth+1$}\;		
		}
		\;
		\SetKwProg{Pn}{Function}{:}{\KwRet}
		\Pn{\FSimulate{$s$, $h$, $depth$}}{
			\uIf{$depth>k$}{\KwRet 0}
			\ElseIf{$depth=k$}{\KwRet $\sum_{s\models\phi_2}b(s,h)$}
			\If{$h\not\in T$}{\For{$a\in A(h,\mathcal{F})$}{$T(ha)\gets (N_{init}(ha),V_{init}(ha),\emptyset)$}
				\KwRet \FRoll{$s$,$h$,$depth$}\;}
			$a\gets\underset{a}{\arg\max}~V(ha)+c\sqrt{\frac{\log {N(h)}}{N(ha)}}$\;
			$(s',z)\sim \mathcal{G}(s,a)$\;
			$R\gets$\FSimulate{$s'$,$haz$,$depth+1$}\;
			$B(h)\gets B(h)\bigcup\{s\}$\;
			$N(h)\gets N(h)+1$\;
			$N(ha)\gets N(ha)+1$\;
			$V(ha)\gets V(ha)+\frac{R-V(ha)}{N(ha)}$\;
			\KwRet $R$		
		}
		\caption{Modified POMCP to check the PCTL specification $\phi=P_{\unlhd p}[\phi_1~\mathcal{U}^{\leq k}\phi_2]$}
		\label{alg::pomcp}
	\end{algorithm}	
\end{figure}

Among different POMDP solvers, we modify a state-of-art POMDP optimal policy computation algorithm, Partially Observable Monte-Carlo Planning (POMCP) \cite{silver2010monte}, that can well fit with our supervisory control framework. POMCP is proposed as an online POMDP planner to find the control policy and optimize a discounted accumulative reward in future. Instead of explicitly solving a POMDP, POMCP applies Monte-Carlo tree search \cite{coulom2006efficient} by running Monte-Carlo simulations to maintain a search tree of histories. Each node in the search tree represents history $h$ as $T(h)=\langle N(h),V(h),B(h)\rangle$. Here $N(h)$ counts the number of times that history $h$ has been visited; $V(h)$ is the value of history $h$; $B(h)$ is a set of particles used to approximate the belief state for history $h$ to avoid exact belief state update for each step. Given the current history $h_t$, each simulation starts in an initial state sampled from the belief state $B(h_t)$. There are two stages of simulation: when the child nodes exist for all children, the actions selection rule follows the Upper Confidence Bounds 1 (UCB1) \cite{auer2002finite} algorithm to maximize $V(ha)+c\sqrt{\frac{\log {N(h)}}{N(ha)}}$ where $c$ is the exploration constant; at the second stage, the actions will be selected following an observation-based rollout policy $\sigma_{rollout}(h,a)$ and normally it follows a uniform random action selection policy. One new node is added to the search tree after each simulation.

To modify POMCP for our model checking purpose for the PCTL specification $\phi=P_{\unlhd p}[\phi_1~\mathcal{U}^{\leq k}\phi_2]$, we use a constant planning depth $k$ instead of a discount factor for the value function to guarantee the termination of each simulation. Meanwhile, without intermediate rewards, a termination reward will be assigned when planning depth $k$ is reached and this reward is equal to $\sum_{s\models\phi_2}b(s,h)$ where $b(\cdot,h)$ is the exact belief state of $h$. For the action selection rules, we limit the selection been considered only on the enabled action set $A(h,\mathcal{F})$ given the supervisor $\mathcal{F}$ and history $h$. While the main algorithm is the same with POMCP, our modified version is shown in Algorithm \ref{alg::pomcp}. Then by initializing the current history $h_0$ to empty, we can estimate the optimal value $V^*(h_0)$, which is equal to the maximum satisfaction probability $V(h_0)$ \cite{zhang2017counterexample}. To find the minimum satisfaction probability, we just need to change the sign of the termination reward, and the estimation is $-V(h_0)$. From the search tree in POMCP, we can also get the selected action for each history node which together gives an observation-based adversary $\sigma$ that can achieve the estimated satisfaction probability. Since our modification on POMCP does not change its main mechanism, the convergence and performance analysis for POMCP is still hold. With the convergence guarantee in probability, the bias of the value function $\mathbb{E}[V(h_0)-V^*(h_0)]$ is $O(\log(N(h_0))/N(h_0)$ \cite{silver2010monte}. Given a fixed $\delta>0$, the probability of $V(h)$ in the range of $[V^*(h)-\Delta_n/n,V^*(h)+\Delta_n/n]$ is less or equal to $\delta$ with $\Delta_n=9\sqrt{2n\ln(2/\delta)}$ for a sufficiently large number of simulations $n$ \cite{kocsis2006bandit}. In practice, we may need to run many simulations (for example, $10^6$) to get a good estimation, but the simulation run can be very fast and the total cost time is still very small (for example, in $10$ to $100$ seconds) as reported in \cite{silver2010monte}. 

\section{Learning based supervisor synthesis }

Within the supervisory control framework using $za$-DFA, our task of finding a supervisor for POMDP is converted to find a DFA, which is an equivalent representation of regular set \cite{kumar2012modeling}. This inspires us to use $L^*$ algorithm to learn a supervisor.

\subsection{$L^*$ Learning Algorithm}

 \begin{table}[!t]
 	\begin{minipage}[b]{0.4\linewidth}
 		\renewcommand{\arraystretch}{1.3}
 		\centering
 		\caption{An Example of the Observation Table in $L^*$ with $\Sigma=\{0,1\}$}
 		\begin{tabular}{c|c}
 			G          & $\epsilon$ \\ \hline
 			$\epsilon$ & 1          \\
 			1          & 0          \\ \hline
 			0          & 1          \\
 			10         & 0          \\
 			11         & 0         
 		\end{tabular}
 		\label{tab::example}
 	\end{minipage}\hfill
 	\begin{minipage}[b]{0.55\linewidth}
 	\centering
 	\begin{tikzpicture}[shorten >=1pt,node distance=3cm,on grid,auto, bend angle=20, thick,scale=0.9, every node/.style={transform shape}]
 	\node[state,initial,accepting] (q_0)   {$q_0$};   
 	\path[->]     
 	(q_0) edge [loop right] node [align=center] {$0$} ();          
 	\end{tikzpicture}
 	
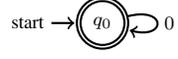
\captionof{figure}{The acceptor DFA corresponding to the observation table in Table \ref{tab::example}}
 	\label{fig::conjecture}	
 	\end{minipage}
 \end{table} 
 
The $L^*$ learning algorithm \cite{angluin1987learning} is proposed to learn an unknown regular set \cite{kumar2012modeling}. Starting from a fixed known size of alphabet $\Sigma$, $L^*$ learning defines an observation table $(Y,E,G)$ to organized the knowledge acquired by the learning algorithm. The row index of the table contains two parts: $Y$ and $Y\cdot\Sigma$, where $Y$ is a nonempty finite prefix-closed set of strings. The column index is given by a nonempty finite suffix-closed set of strings $E$. The function $G$ maps a string $y\in\Sigma^*$ to $\{0,1\}$ where $\Sigma^*$ is the set of all finite length strings containing symbols from $\Sigma$. For a string $y\in ((Y\cup Y\cdot\Sigma)\cdot E)$, $G(y)=1$ if and only if $y\in U$. For each row entry of a string $y$, its row denotes the finite function $f$ from $E$ to $\{0,1\}$ defined by $f(e)=G(y\cdot e)$. Initializing the observation table with $Y=E=\{\epsilon\}$, $L^*$ algorithm tries to make the table \textsl{closed} and \textsl{consistent}. For closeness, $\forall y_1 \in Y\cdot\Sigma$, it requires that $\exists y_2\in Y$, s.t. $row(y_1)=row(y_2)$; for consistence, whenever $y_1,y_2\in Y$ with $row(y_1)=row(y_2)$, it requires that $\forall \alpha\in\Sigma$, $row(y_1\cdot \alpha)=row(y_2\cdot \alpha)$ \cite{angluin1987learning}. Given a closed and consistent observation table, a DFA $\mathcal{F}=\{Q,\bar{q},\Sigma,\delta,Q_m\}$ as the acceptor can be generated with its accepting language $L_m(\mathcal{F})$ representing the learned regular set as follows: 
\begin{itemize}
	\item $Q=\{row(y):y\in Y\}$,
	\item $\bar{q}=row(\epsilon)$,
	\item $\sigma(row(y),\alpha)=row(y\cdot \alpha)$.
	\item $Q_m=\{row(y):y\in Y\text{ and }G(y)=1\},$
\end{itemize} 
For the observation table shown in Table \ref{tab::example}, it is closed and consistent, and the corresponding DFA is shown in Fig. \ref{fig::conjecture}.

To generate a closed and consistent observation table, $L^*$ learning maintains a Questions \& Answers mechanism. Given the alphabet $\Sigma$, two types of questions, membership query and conjecture, are asked by the \textsl{Learner} and answered by the \textsl{Teacher}. For the membership query, the Learner asks whether a string $y\in\Sigma^*$ is a member of $U$ or not, and the Teacher answers $true$ or $false$, respectively. For the conjecture, the Learner asks whether a learned regular set is equal to $U$ or not, and the Teacher answers $true$, or $false$ with a string $y_c$ showing the symmetric difference between the learned set and $U$. In the latter case, $y_c$ is called a counterexample. With the membership query, if the table is not closed, the algorithm finds $y_1\in Y$, $\alpha\in \Sigma$ s.t., $row(y_1\cdot \alpha)\neq row(y_2),\forall y_2\in Y$, then adds $y_1\cdot \alpha$ to $Y$ and extends the table; if the table is not consistent, the algorithm finds $y_1,y_2\in Y, \alpha\in \Sigma, e\in E$, s.t., $row(y_1)=row(y_2)$ but $G(y_1\cdot \alpha\cdot e)\neq G(y_2\cdot \alpha\cdot e)$, then adds $\alpha\cdot e$ to $E$ and extends the table \cite{angluin1987learning}. With the conjecture, if $y_c$ is given as a counterexample, $y_c$ and its prefixes will be added to $Y$ and the table is extended using membership queries. With a Teacher being able to answer membership queries and conjectures, $L^*$ algorithm is proved to converge to the minimum DFA accepting $U$ in polynomial time \cite{angluin1987learning}. 

\subsection{Learn za-DFA as the Supervisor}

\begin{figure*}[!t]
	\centering
	\includegraphics[scale=0.75]{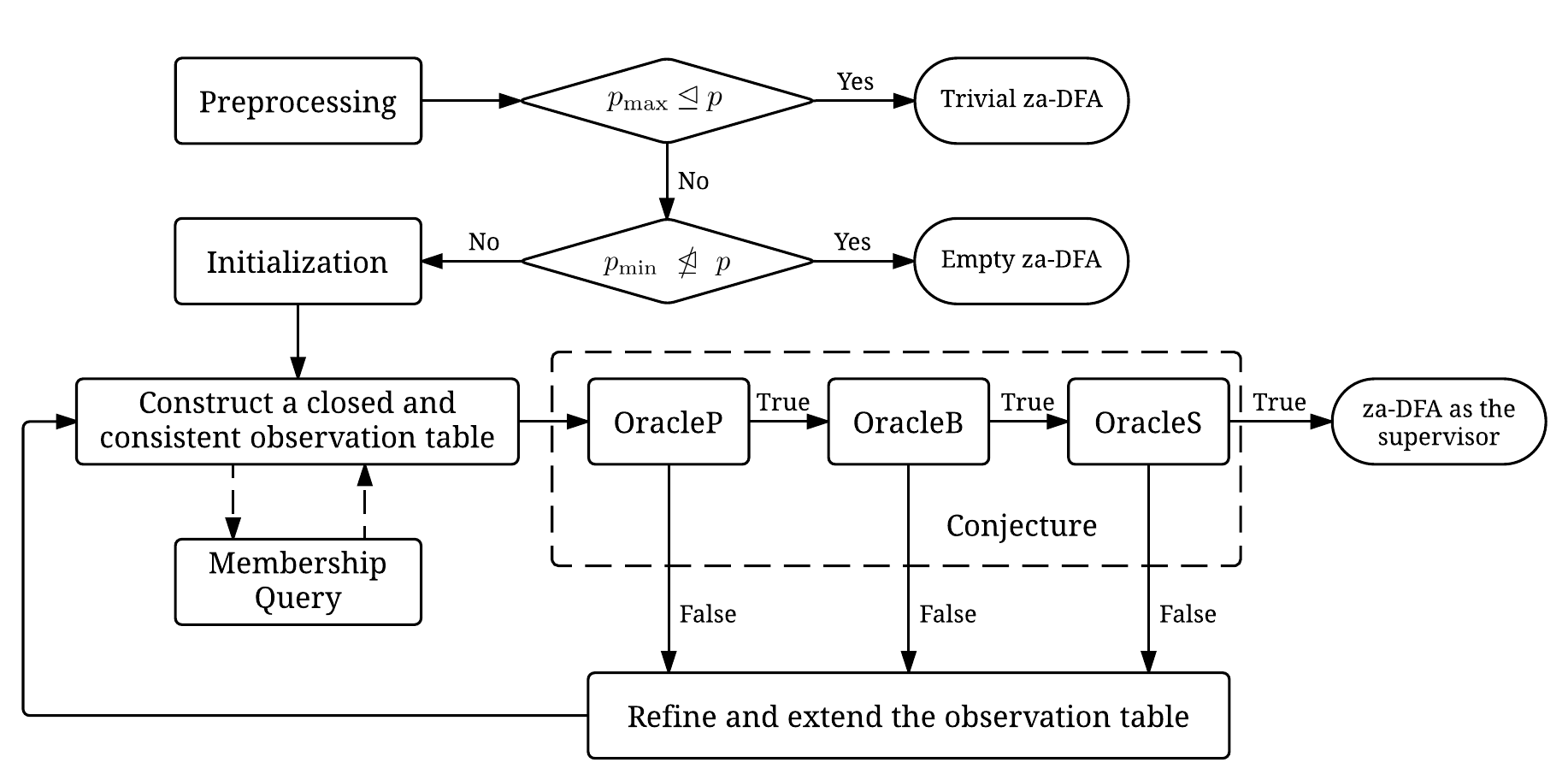}
	\caption{$L^*$ learning based supervisor synthesis of POMDP}
	\label{figure_flowchart}
\end{figure*}

Given a POMDP $\mathcal{P}=\{S,\bar{s},A,Z,T,O\}$ and a finite horizon PCTL specification $\phi$, we use $L^*$ learning to learn a za-DFA $\mathcal{F}$ as the supervisor. To get a feasible za-DFA that can regulate POMDP to satisfy the specification $\phi$, we develop algorithms to answer membership queries and conjectures. To simplify the analysis, we will take $\phi=P_{\unlhd}[\phi_1 \mathcal{U}^k\phi_2]$ with $\unlhd\in\{\leq,<\}$ as the specification to illustrate the learning process. The overview of the learning process is shown in Fig. \ref{figure_flowchart} and we illustrate it as follows.

\subsubsection{Preprocessing}

Before the initialization of $L^*$ learning, we first find the observation-based adversaries $\sigma_{max}$ and $\sigma_{min}$ that give the maximum and minimum satisfaction probabilities $p_{\max}$ and $p_{\min}$ for the path formula $\psi=\phi_1 \mathcal{U}^k\phi_2$, respectively. With the probability bound in $\phi$ given by $\unlhd$, we compare $p_{\max}$ and $p_{\min}$ with the threshold $p$: if $p_{\max}\unlhd p$ then any observation-based adversaries can be applied and a trivial za-DFA with one state and self loop transitions under any $\langle z,a\rangle\in \Sigma$ will be returned as the supervisor; if $p_{\min}\ntrianglelefteq p$ then no observation-based adversaries can be applied and an empty za-DFA that only accepts the empty string $\epsilon$ will be returned. 

\subsubsection{Initialization} 

After the preprocessing stage to calculate $\sigma_{max}$, $\sigma_{min}$ and their corresponding $p_{\max}$, $p_{\min}$, we can initialize the $L^*$ learning algorithm. Starting with the alphabet $\Sigma$ defined in Definition \ref{zaDFA}, the observation table $(Y,E,G)$ is initialized with $Y=\{\epsilon\}$, $E=\{\epsilon\}$ and $G(\epsilon)=1$. Then membership queries are generated by the Learner to extend the table.

Beside the observable table, we initialize two string sets $C_B$ and $C_S$ to empty. Here $C_B$ and $C_S$ will contain strings of negative counterexamples returned from the \textit{OracleB} and \textit{OracleS}, respectively, and both oracles will be introduced in the conjecture answering section.  

\subsubsection{Answering Membership Queries}

For each string $y=\alpha_0\alpha_1...\alpha_n,\alpha_i\in \Sigma$, the membership query checks whether or not the corresponding observation-action sequence $obs=\langle z_0,a_0\rangle\langle z_1,a_1\rangle...\langle z_n,a_n\rangle$ can be used as the control policy for histories as the prefix of $obs$. If there exists a prefix of $y$ in $C_S\cup C_B$, the membership query returns $false$. Otherwise, we will unfold the POMDP $\mathcal{P}$ given the control policy from $y$. This unfolding process follows the product MDP generation rules given in Definition \ref{def::Mza}. Basically $y$ can be converted to a za-DFA with a unique action being selected for a history as the prefix of $y$. Then its product MDP $\mathcal{M}^{\mathcal{F}}$ turns into a DTMC. On DTMC $\mathcal{M}^{\mathcal{F}}$, the model checking result of specification $\phi$ will answer the membership query with $true$ if and only if $\mathcal{M}^{\mathcal{F}}\models\phi$. If $|y|>k$, we will take its prefix $y':|y'|=k$ and apply the membership query for $y'$ since the specification only constrain the regulated behavior up to the depth $k$.    

\noindent\textbf{Remark}: In the original $L^*$ algorithm, $G(y)=1$ implies that the unknown regular set $U$ accepts $y$. But in our case, if membership query returns $true$, it only means the corresponding control policy will not cause the violation of the specification by itself. Here $y$ may still need to be removed to get a correct supervisor because the satisfaction probability of the regulated behavior is determined based on the accumulative probability brought by different strings accepted in the supervisor. 

Based on the $L^*$ algorithm, the Learner will keep generating membership queries until a closed and consistent table $(Y,E,G)$ is learned. Then a za-DFA $\mathcal{F}$ is generated as the acceptor of $(Y,E,G)$.  

\subsubsection{Answering Conjectures}
Given a za-DFA $\mathcal{F}$ as the acceptor, the Learner asks a conjecture to check whether or not $\mathcal{F}$ is a non-blocking supervisor that can regulate POMDP $\mathcal{P}$ to satisfy PCTL $\phi$. If the answer is $true$, the algorithm will terminate with the learned za-DFA as a non-blocking and permissive supervisor. Otherwise, counterexamples will be returned to guide the refinement and extension process of the observation table for the Learner. To answer conjectures, three oracles are defined to guarantee the soundness and completeness of our learning algorithm.

Since we know $\sigma_{\min}$ is a suitable adversary that will not violate the specification or cause blocking during the POMDP execution, we define \textit{OracleP} to check whether or not there exists a string $y\in \Sigma^*$ such that $y=obs(\rho)\cdot a$ with $\sigma_{\min}(obs(\rho))=a$ but $y\not\in L_m(\mathcal{F})$. If yes, the conjecture will answer $false$ with $y$ being returned as a positive counterexample to make $y\in L_m(\mathcal{F})$. With OracleP, we can guarantee that the learned supervisor $\mathcal{F}$ will accept any history-action pairs given by $\sigma_{\min}$.  

\noindent\textbf{Remark}: For any string $y=obs(\rho)\cdot a$ with $\sigma_{\min}(obs(\rho))=a$, the membership query will return $true$. That is because if this single control policy could bring a probability violating the requirement in the specification, $p_{\min}$ brought by $\sigma_\{min\}$ will also violate the requirement, which will terminate the algorithm in the preprocessing stage. Therefore, there are no conflicts between the membership query and OracleP.    

If OracleP does not find a positive counterexample, we use OracleB to check whether or not $\mathcal{F}$ is non-blocking. Here we checks all states $szq\in S^{\mathcal{F}}$ that are $k$-step reachable from $\bar{s}\bar{q}$ on $\mathcal{M}^{\mathcal{F}}$: if all such states have outgoing transitions being defined, OracleB returns $true$; otherwise, OracleB will check the causes of blocking. Assume there exists a $k$-step reachable state $szq$ that does not have any outgoing transitions. Then $q\in Q_m$ and by applying a depth-first search on $\mathcal{M}^{\mathcal{F}}$ we can find the shortest observation-action sequence transits from $\bar{s}\bar{q}$ to $szq$. Denote this string as $y$. Depending on whether $y\in Y$ or not, we have two possible causes for the blocking on $szq$. If $y\not\in Y$, the blocking of the supervisor is because the Learner can generate a conjecture without adding $y$ to $Y$ and asking membership queries for $y\cdot \Sigma$. To know if there exists an action that can be enabled for $szq$ to fix the blocking, OracleB will return $false$ with $y$ as the counterexample to enforce $y\cdot\Sigma$ appearing as rows in the observation table. If $y$ has already been included in $Y$, $y\cdot\alpha$ for all $\alpha\in\Sigma$ will appear as rows in the observation table. Since $q$ has no outgoing transitions under the observation $z$, all strings $y\cdot\alpha$ with $\alpha=\langle z,\cdot\rangle$ have been answered by membership queries with $false$. This means once the POMDP execution reaches state $s$ and observes $z$ while the za-DFA reaches state $q$, choosing any action will cause the violation of the specification under current za-DFA. Therefore, $szq$ should be avoided during the system transition. To remove the strings that may lead to such states, all states $szq\in S^{\mathcal{F}}$ with no outgoing transitions will be marked as dark states and the transitions to dark states will be removed in $\mathcal{M}^{\mathcal{F}}$. This process keeps running until no new dark state appears. From traces starting from the initial state $\bar{s}\bar{q}$ to dark states, the observation-action sequences are extracted and form a string set $C_b$. Then $C_B$ is updated to $C_B=C_B\cup C_b$. OracleB will return $false$ together with the shortest string $y_c\in C_B$ as the negative counterexample.

If OracleB does not find a counterexample, we use OracleS to check whether or not $\mathcal{M}^{\mathcal{F}}\models_{obs}\phi$. If no, OracleS will return $false$ with a negative counterexample $y_c$ as the evidence of specification violation. To find such a string $y_c$ as the counterexample, we first solve $\mathcal{M}^{\mathcal{F}}$ and find the observation-based adversary $\sigma_c$ that gives the maximum satisfaction probability $p_c:~p_c \ntrianglelefteq p$. With $\sigma_c$, we generate a derived DTMC with histories as states: $\tilde{\mathcal{M}}=\{\mathcal{H}\cup \{h_d\},h_0,\tilde{T}\}$. Here $\mathcal{H}$ is the state space of histories with $h_0=empty$ and $h_d$ is a dummy state. For $h\in \mathcal{H}$ with $|h|<k$, $\tilde{T}(h,haz)=\sum_{s\in S}\sum_{s'\in S}b(s,h)T^*(s,a,s')O(s',z)$ with $b(s,h)$ the belief state function, $\sigma_c(h)=a$ and 
\begin{align*}
	T^*(s,a,s')= 
	\begin{cases}
	\delta_{s}(s'),~s\models\neg\phi_1 \vee\phi_2;\\
	T(s,a,s'),\text{ otherwise,}
	\end{cases}	
\end{align*}
where $\delta_{s}$ is the standard Dirac delta function to make $s\models\neg\phi_1 \vee\phi_2$ absorb. For $h\in \mathcal{H}$ with $|h|=k$, $\tilde{T}(h,h_d)=\sum_{s\models \phi_2}b(s,h)$. Basically we are grouping up paths with the same observation sequences together in $\mathcal{M}^{\mathcal{F}|\sigma_c}$ and generate $\tilde{\mathcal{M}}$. Therefore, a path in $\tilde{\mathcal{M}}$ corresponds to a set of paths in $\mathcal{M}^{\mathcal{F}|\sigma_c}$. For a path in $\tilde{\mathcal{M}}$ that starts from $h_0$ and ends in $h_d$, its transition probability is equal to the accumulative transition probability of the corresponding set of paths in $\mathcal{M}^{\mathcal{F}|\sigma_c}$ ending in a state with label $\phi_2$ in $k$ steps. Since $\sigma_c$ witnesses the violation of the specification, $\tilde{\mathcal{M}}\not\models P_{\unlhd p}[true~\mathcal{U}^{k+1}h_d]$. Then we apply the DTMC counterexample generation algorithm in \cite{han2009counterexample} to get the strongest evidence as a finite path with the maximum probability of violate. Denote its corresponding observation-action sequence as $y_c$. If $y_c=obs(\rho)\cdot a$ while $\sigma_{\min}(obs(\rho))=a$, $y_c$ will be replaced by the observation-action sequence of the path with the second largest probability of violation. This process keeps going until $y_c$ does not conflict with $\sigma_{\min}$. Then $y_c$ will be returned as the negative counterexample and $C_S$ is updated with $C_S=C_S\cup \{y_c\}$.

If all three oracles return $true$, our algorithm will return the result $za$-DFA as the supervisor and terminate. If there exist counterexamples returned from either oracle, the observation table will be refined and extended.

\subsubsection{Refining and Extending the Observation Table}

In the next iteration, given a counterexample $y$ returned from conjectures and the updated $C_B$ and $C_S$, we first refine the observation table by correcting $G(y')=1$ to $G(y')=0$ if $Pref(y')\in C_B\cup C_S$. Then $y$ and all its prefixes are added to $Y$ in $(Y,E,G)$. After that, the observation table is extended using membership queries to generate a new closed and consistent table.

\section{Analysis and Discussions}

We analyze the $L^*$ learning based supervisor synthesis algorithm in this section regards to the termination, soundness, and completeness, as well as the computational complexity. Our analysis focuses on the cases where the algorithm is not terminated during the preprocessing stage since trivial statements can be followed otherwise. 

\subsection{Termination}

In the $L^*$ learning, we use membership queries and conjectures to collect information about whether or not an observation-action sequence can be used as part of a proper supervisor. Because we consider finite POMDP with a finite horizon specification, the number of all possible observation-action sequences are finite. So we only have a finite number of strings needed to be labeled in the observation table for the $L^*$ algorithm. Our algorithm requires a refinement process for the observation table if the returned negative counterexamples and their suffixes were answered with $true$ by membership queries in previous iterations. However, for a string $y\in (Y\cup Y\cdot\Sigma)\cdot E$, it will never happen that $G(y)$ is changed from $0$ to $1$. Consider a string $y$ with $G(y)=0$. Then either the accumulative probability from $y$ violates the threshold given by the specification, or $Pref(y)\in C_B\cup C_S$. In any of these cases, membership queries will always return $false$ for $y$. While OracleP will return certain strings as positive counterexamples, $y$ will never be returned by OracleP, i.e., $y$ is not accepted by $\sigma_{\min}$. If the accumulative probability from $y$ violates the threshold, it will never belong to $\sigma_{\min}$. If $y\in C_S$, by definition of OracleS, $y$ cannot be returned by OracleP. If $y\in C_B$, $y$ must be returned by OracleB which will only happen after OracleP returns $true$. But when OracleP returns $true$, the observation-action sequences from $\sigma_{\min}$ are all accepted by the acceptor za-DFA, and none of them will cause blocking of the supervisor which is guaranteed by $\sigma_{\min}$ as an observation-based adversary. Therefore, if $y\in C_B$, $y$ can never belong to $\sigma_{\min}$. As a result, $G(y)$ will never be changed from $0$ to $1$. With the fact that the number of strings to be inquired is finite and at each iteration the algorithm must return counterexamples if any oracles return $false$, we can conclude that the termination of our supervisor synthesis algorithm is guaranteed. The upper bound of the number of iterations is equal to twice of the number of possible strings.

\subsection{Soundness and Completeness}

Our $L^*$ learning based supervisor synthesis algorithm is sound and complete.  If a za-DFA is returned as the supervisor, based on the definition of OracleP, OracleB, and OracleS, this za-DFA is non-blocking, and the model checking on the regulated behavior of POMDP proves the satisfaction of the specification. This shows the soundness of the algorithm.

For the completeness, if there exists a proper supervisor, our algorithm will return a za-DFA representing $\sigma_{\min}$ in the worst cases. This is guaranteed by OracleP. But we cannot guarantee the permissiveness for the worse cases when OracleS returns "good" observation-action sequences as negative counterexamples. While OracleS will never misidentify a single string carrying enough probability mass of violation, if a set of paths is needed to witness the violation, how to select a proper counterexample from that set is still a research question, and it is possible that some paths accepted by the desired supervisor are returned as negative counterexamples. While now we will return the one with the maximum probability mass, newly developed counterexample selection algorithms for probabilistic systems can be applied and improve the performance of our learning framework.  

\subsection{Complexity}
Define the size of the POMDP $S_{\mathcal{P}}$ as the product of the size of the underlying MDP $S_{\mathcal{M}}$ and $|Z|$: $S_{\mathcal{P}}=S_{\mathcal{M}}*|Z|$ and denote the planning horizon of the specification as $k$. Then following the termination analysis, the number of iterations is at most $O(|\Sigma|^k)$ where $\Sigma$ is the alphabet. In each iteration, denote the size of current acceptor DFA as $S_{\mathcal{F}}$. OracleP tries to find the difference between the current acceptor DFA and $\sigma_{\min}$. This can be achieved with time complexity $O(S_{\mathcal{F}})$ by doing complement and interaction between two DFAs then applying depth first search to check whether or not the initial state can be reached in $k$ steps from the accepted state. OracleB mainly applies depth first search on the product MDP, so the time complexity is $O(S_{\mathcal{P}}*S_{\mathcal{F}})$. OracleS replies on POMDP solving which generally have a time complexity exponential with $k$, linear with the length of the PCTL formula $L_f$ (the number of logical and temporal operators in the formula). But with the modified POMCP method, the model checking result can be returned in seconds by running thousands of simulations and the running time will depend on the hardware. After that, the counterexample selection algorithm will take polynomial time with $k$ and the number of transitions in the derived DTMC \cite{han2009counterexample}. In the learning process, the maximum number of membership queries is at most $O\left(k|\Sigma|^k\right)$. Then combining with the time analysis of $L^*$ in \cite{angluin1987learning}, we can see that our algorithm has a complexity exponential with $k$, polynomial with $S_{\mathcal{P}}$, and $L_f$. However, whenever we eliminate negative counterexamples, their suffixes are also removed. Therefore the $S_{\mathcal{F}}$ and the number of iterations rarely assume large values in practice. So this complexity analysis is rather conservative. 

\section{Example}

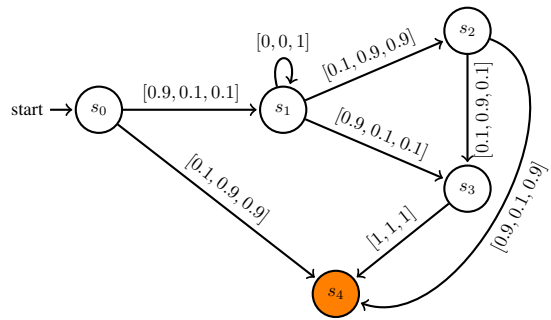
\begin{figure}[!t]
	\centering	
	\begin{tikzpicture}[shorten >=1pt,node distance=3.5cm,on grid,auto, bend angle=20, thick,scale=0.7, every node/.style={transform shape}] 
	\node[state,initial] (s0) {$s_0$};
	\node[state] (s1) [right=3.5cm of s0] {$s_1$};
	\node[state] (s2) [above right=1.5cm and 3.5cm of s1] {$s_{2}$};    		
	\node[state] (s3) [below right=1.5cm and 3.5cm of s1] {$s_{3}$};  	
	\node[state, fill=orange] (s4) [below right=3.5cm and 1cm of s1] {$s_{4}$};
	
	\path[->]
	(s0) edge node [pos=0.5, sloped, above]{$[0.9,0.1,0.1]$} (s1)  
	edge node[pos=0.5, sloped, above] {$[0.1,0.9,0.9]$} (s4) 	
	
	(s1) edge  node [pos=0.5, sloped, above]{$[0.1,0.9,0.9]$} (s2) 
	edge  node  [pos=0.5, sloped, above] {$[0.9,0.1,0.1]$} (s3) 
	edge [loop above, pos = 0.5,sloped, above] node {$[0,0,1]$} ()		
	(s3) edge  [pos=0.5, sloped, above] node{$[1,1,1]$} (s4)
	(s2) edge [out=340, in=340,pos=0.5, sloped, below] node{$[0.9,0.1,0.9]$} (s4)
	edge  [pos=0.5, sloped, below] node{$[0.1,0.9,0.1]$} (s3)
	
	; 
	\end{tikzpicture} 
	\caption{The POMDP model $\mathcal{P}$}
	\label{fig:P}
\end{figure}

Consider a POMDP $\mathcal{P}=\{S,\bar{s},A,Z,T,O\}$, where
\begin{itemize}
	\item $S= \{s_0,s_1,s_2,s_3,s_4\}$;
	\item $\bar{s} = s_0$;	
	\item $A=\{a_1,a_2,a_3\}$;
	\item $Z=\{z_1,z_2\}$.
\end{itemize}
The transition probabilities under different actions are given in the order of $a_1$, $a_2$, $a_3$ in the square brackets shown in Fig. \ref{fig:P}. The observation matrix is given in Table \ref{table:O}. Among $S$, the state $s_4$ represents a failure state with label $fail$ and is colored by orange in Fig. \ref{fig:P}. The specification is given by a finite horizon PCTL $\phi=\mathcal{P}_{\leq 0.28}[\psi]$ with $\psi=true~\mathcal{U}^{\leq 3}fail$, which requires the probability of reaching failure within $3$ steps should be less or equal to $0.28$.

\noindent\textbf{Remark}: This POMDP $\mathcal{P}$ is specially designed that the model checking problem can be solved quite straightforwardly. Then we can focus on the illustration of our supervisor synthesis algorithm.

\begin{table}[!t]
	\centering
	\makebox[0pt][c]{\parbox{\linewidth}{
			\begin{minipage}[t]{.5\linewidth}
				\centering
				\caption{The Observation matrix of POMDP $\mathcal{P}$}
				\label{table:O}
				\begin{tabular}{c|cc}
					$O(s,z)$ & $z_1$ & $z_2$ \\ \hline
					$s_0$    & 0.3   & 0.7   \\
					$s_1$    & 0.5   & 0.5   \\
					$s_2$    & 0.2   & 0.8   \\
					$s_3$    & 1   & 0   \\
					$s_4$    & 0     & 1    
				\end{tabular}
				\bigskip
				\centering
				\caption{The closed and consistent observation table in the first iteration}
				\label{tab::G1}
				\begin{tabular}{c|c}
					$G$        & $\epsilon$ \\ \hline
					$\epsilon$ & 1          \\ 
					2          & 0          \\	\hline
					1          & 1          \\		
					3          & 0          \\
					4          & 1          \\
					5          & 0          \\
					6          & 0          \\
					2$\{1,...,6\}$          & 0          \\
				\end{tabular}
				
			\end{minipage}
			\hfil
			\begin{minipage}[t]{.5\linewidth}
				\centering
				\caption{The closed and consistent observation table in the second iteration}
				\label{tab::G2}
				\begin{tabular}{c|ccc}
					G              & $\epsilon$ & 3 & 13 \\ \hline
					$\epsilon$     & 1          & 0 & 1  \\
					2              & 0          & 0 & 0  \\
					1              & 1          & 1 & 0  \\
					13             & 1          & 1 & 1  \\
					11             & 1          & 0 & 0  \\ \hline
					3              & 0          & 0 & 0  \\
					4              & 1          & 1 & 0  \\
					5              & 0          & 0 & 0  \\
					6              & 0          & 0 & 0  \\
					2$\{1,...,6\}$ & 0          & 0 & 0  \\
					12             & 1          & 1 & 1  \\
					14             & 1          & 0 & 0  \\
					15             & 1          & 1 & 1  \\
					16             & 1          & 1 & 1  \\
					13$\{1...,6\}$ & 1          & 1 & 1  \\
					11$\{1,2,3\}$  & 0          & 0 & 0  \\
					11$\{4,5,6\}$  & 1          & 1 & 1 
				\end{tabular}
			\end{minipage}
		}}
	\end{table}

	\begin{figure}[!t]
		\centering
		\begin{minipage}[t]{.42\linewidth}
			\centering
			\begin{tikzpicture}[shorten >=1pt,node distance=2cm,on grid,auto, bend angle=20, thick,scale=0.7, every node/.style={transform shape}]
			\node[state,initial,accepting] (q0)   {$q_0$};  
			\node[state,accepting] (q1) [right=2.5cm of q0] {$q_1$}; 
			\path[->]     
			(q0) 
			edge node[pos=0.5, sloped, above] {$1,4$} (q1)
			
			(q1) edge [loop above] node [align=center] {$3,6$} () ;	
			\end{tikzpicture}
			\centering\caption{The za-DFA $\mathcal{F}_{\min}$ that gives the minimum probability of satisfying $\psi$}\label{fig::Fmin}
		\end{minipage}
		\hfil
		\begin{minipage}[t]{.42\linewidth}
			\centering	
			\begin{tikzpicture}[shorten >=1pt,node distance=2cm,on grid,auto, bend angle=20, thick,scale=0.7, every node/.style={transform shape}]
			\node[state,initial,accepting] (q0)   {$q_0$};  
			\node[state] (q1) [right=2.5cm of q0] {$q_1$}; 
			\path[->]     
			(q0) edge [loop above] node [align=center] {$1,4$} ()
			edge node[pos=0.5, sloped, above] {$2,3,5,6$} (q1) ;	
			\end{tikzpicture}
			\centering\caption{The za-DFA $\mathcal{F}_{1}$ as the acceptor in the first iteration}
			\label{fig::F1}
		\end{minipage}
	\end{figure}
	
	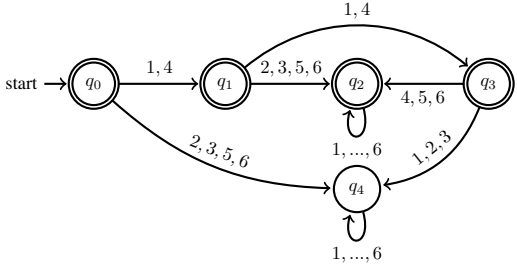
\begin{figure}[!t]
		\centering	
		\begin{tikzpicture}[shorten >=1pt,node distance=2cm,on grid,auto, bend angle=20, thick,scale=0.7, every node/.style={transform shape}]
		\node[state,initial,accepting] (q0)   {$q_0$};  
		\node[state, accepting] (q1) [right=2.5cm of q0] {$q_1$}; 
		\node[state, accepting] (q2) [right=2.5cm of q1] {$q_2$}; 
		\node[state, accepting] (q3) [right=2.5cm of q2] {$q_3$}; 
		\node[state] (q4) [below=2cm of q2] {$q_4$}; 
		\path[->]     
		(q0) 
		edge node[pos=0.5, sloped, above] {$1,4$} (q1) 
		edge [bend right = 20]  node[pos=0.5, sloped, above] {$2,3,5,6$} (q4) 
		(q1) edge node [align=center] {$2,3,5,6$} (q2)	
		edge[bend left = 40]  node[ pos=0.5, sloped, above] {$1,4$} (q3) 
		(q2) edge [loop below] node [align=center] {$1,...,6$} ()
		(q3) edge node [align=center] {$4,5,6$} (q2)	
		edge[bend left = 30]  node[ pos=0.5, sloped, above] {$1,2,3$} (q4)
		(q4) edge [loop below] node [align=center] {$1,...,6$} ()
		
		;
		\end{tikzpicture}
		\centering\caption{The za-DFA $\mathcal{F}_{2}$ as the acceptor in the second iteration}
		\label{fig::F2}
	\end{figure}

Based on the observation and action sets, we have the alphabet $\Sigma=\{1,2,3,4,5,6\}$ for the za-DFA as the supervisor where
\begin{itemize}
	\item $1:\langle z_1,a_1\rangle$, $2:\langle z_1,a_2\rangle$, $3:\langle z_1,a_3\rangle$,
	\item $4:\langle z_2,a_1\rangle$, $5:\langle z_2,a_2\rangle$, $6:\langle z_2,a_3\rangle$.
\end{itemize} 

During the preprocessing stage, it is not hard to see that the minimum probability of satisfying $\psi$ is $0$ given by the adversary shown as the za-DFA $\mathcal{F}_{\min}$ in Fig. \ref{fig::Fmin} while the maximum probability of satisfying $\psi$ is $0.96$. Since $p_{\min}<0.28<p_{\max}$, we are ready to initialize the $L^*$ algorithm. 

In the first iteration, after answering membership queries, we construct a closed and consistent observation table shown in Table \ref{tab::G1} with the corresponding acceptor za-DFA $\mathcal{F}_1$ shown in Fig. \ref{fig::F1}. Since $G(3)=0$ and $G(6)=0$, the acceptor is generated with $3,6 \not\in L_m(\mathcal{F}_1)$. This makes OracleP return $false$ with $13$ as the positive counterexample. 	

In the second iteration, after adding $13$ to $Y$ and extending the table with membership queries, we have a new closed and consistent observation table shown in Table \ref{tab::G2}. While OracleP returns $true$, OracleB finds that at state $s_3z_1q_3$ no outgoing transitions is defined since $\delta(q_3,\{1,2,3\})=q_4$ while $q_4\not\in Q_m$ in $\mathcal{F}_2$. Then string $11$ is selected as the shortest observation-action sequence transiting to $s_3z_1q_3$. Because $11\in Y$ in the observation table, OracleB will mark dark states and disable transitions leading to the dark states. Then the string set $C_b=\{11,14,41,44\}$ is returned and $C_B=\{11,14,41,44\}$. With $11$ as the negative counterexample and updated $C_B$, we need to refine the table.

\begin{figure}[!t]
	\centering	
	\begin{tikzpicture}[shorten >=1pt,node distance=2cm,on grid,auto, bend angle=20, thick,scale=0.7, every node/.style={transform shape}]
	\node[state,initial,accepting] (q0)   {$q_0$};  
	\node[state, accepting] (q1) [right=2cm of q0] {$q_1$}; 
	\node[state, accepting] (q2) [right=2cm of q1] {$q_2$}; 
		\node[state, accepting] (q3) [right=2cm of q2] {$q_3$}; 

	\path[->]     
	(q0) 
	edge node[pos=0.5, sloped, above] {$1,4$} (q1) 

	(q1) edge node [align=center] {$2,5$} (q2)	

	(q2) edge node [align=center] {$1,4$} (q3)	
	;	
	\end{tikzpicture}
	\centering\caption{The witness za-DFA $\mathcal{F}_{c}$ in the third iteration}
	\label{fig::Fc3}
\end{figure}
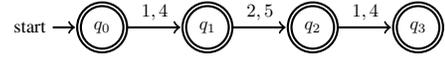

In the third iteration, we first correct and extend the observation table to a closed and consistent one shown in Table \ref{tab::G3}. Now both OracleP and OracleB return $true$. But OracleS finds an adversary $\sigma_c$ shown as the za-DFA $\mathcal{F}_c$ in Fig. \ref{fig::Fc3} gives the maximum probability $0.919$ of satisfying $\psi$. In the derived DTMC $\tilde{\mathcal{M}}$, with the largest probability $0.2916$ of satisfying $true~\mathcal{U}^{\leq 4}h_d$, $124$ is returned as the negative counterexample and $C_S=\{124\}$.

\begin{table}[!t]
	\centering
	\caption{The observation table in the third iteration}
	\label{tab::G3}
\begin{tabular}{c|ccc}
	G              & $\epsilon$                        & 3                                 & 13                                \\ \hline
	$\epsilon$     & 1                                 & 0                                 & 1                                 \\
	2              & 0                                 & 0                                 & 0                                 \\
	1              & 1                                 & 1                                 & 0                                 \\
	13             & 1                                 & 1                                 & 1                                 \\
	11             & {\color[HTML]{FE0000} $\cancel{1}\rightarrow 0$} & 0                                 & 0                                 \\ \hline
	3              & 0                                 & 0                                 & 0                                 \\
	4              & 1                                 & 1                                 & 0                                 \\
	5              & 0                                 & 0                                 & 0                                 \\
	6              & 0                                 & 0                                 & 0                                 \\
	2$\{1,...,6\}$ & 0                                 & 0                                 & 0                                 \\
12             & 1                                 & 1                                 & 1                                 \\
14             & {\color[HTML]{FE0000} $\cancel{1}\rightarrow 0$}& 0                                 & 0                                 \\
15             & 1                                 & 1                                 & 1                                 \\
16             & 1                                 & 1                                 & 1                                 \\
13$\{1...,6\}$ & 1                                 & 1                                 & 1                                 \\
11$\{1,2,3\}$  & 0                                 & 0                                 & 0                                 \\
11$\{4,5,6\}$  & {\color[HTML]{FE0000} $\cancel{1}\rightarrow 0$} & {\color[HTML]{FE0000} $\cancel{1}\rightarrow 0$} & {\color[HTML]{FE0000} $\cancel{1}\rightarrow 0$}
\end{tabular}
\end{table}

\begin{figure}[!t]
	\centering	
	\begin{tikzpicture}[shorten >=1pt,node distance=2cm,on grid,auto, bend angle=20, thick,scale=0.7, every node/.style={transform shape}]
	\node[state,initial,accepting] (q0)   {$q_0$};  
	\node[state, accepting] (q1) [right=2.5cm of q0] {$q_1$}; 
	\node[state, accepting] (q2) [right=2.5cm of q1] {$q_2$}; 

	\node[state] (q4) [below=1.8cm of q1] {$q_3$}; 
	\path[->]     
	(q0) 
	edge node[pos=0.5, sloped, above] {$1,4$} (q1) 
	edge [bend right = 20]  node[pos=0.5, sloped, above] {$2,3,5,6$} (q4) 
	(q1) edge node [align=center] {$2,3,5,6$} (q2)	
edge node [align=center] {$1,4$} (q4)
	(q2) edge [loop right] node [align=center] {$1,...,6$} ()

	(q4) edge [loop right] node [align=center] {$1,...,6$} ()
	
	;	
	\end{tikzpicture}
	\centering\caption{The za-DFA $\mathcal{F}_{3}$ as the acceptor in the third iteration}
	\label{fig::F3}
\end{figure}
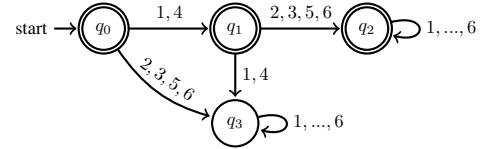

\begin{figure}[!t]
	\centering	
\begin{tikzpicture}[shorten >=1pt,node distance=2cm,on grid,auto, bend angle=20, thick,scale=0.7, every node/.style={transform shape}]
\node[state,initial,accepting] (q0)   {$q_0$};  
\node[state, accepting] (q1) [right=2.5cm of q0] {$q_1$}; 
\node[state, accepting] (q2) [right=2.5cm of q1] {$q_2$}; 
\node[state, accepting] (q3) [right=2.5cm of q2] {$q_3$}; 
\node[state] (q4) [below=2cm of q2] {$q_4$}; 
\path[->]     
(q0) 
edge node[pos=0.5, sloped, above] {$1,4$} (q1) 
edge [bend right = 20]  node[pos=0.5, sloped, above] {$2,3,5,6$} (q4) 
(q1) edge node [align=center] {$3,6$} (q2)	
edge[bend left = 30]  node[ pos=0.5, sloped, above] {$2,5$} (q3) 
(q2) edge [loop below] node [align=center] {$1,...,6$} ()
(q3) edge node [align=center] {$1,2,3,5$} (q2)	
edge[bend left = 30]  node[ pos=0.5, sloped, above] {$4,6$} (q4)
(q4) edge [loop below] node [align=center] {$1,...,6$} ()

;
\end{tikzpicture}
	\centering\caption{The za-DFA $\mathcal{F}_{4}$ as the acceptor in the fourth iteration}
	\label{fig::F4}
\end{figure}
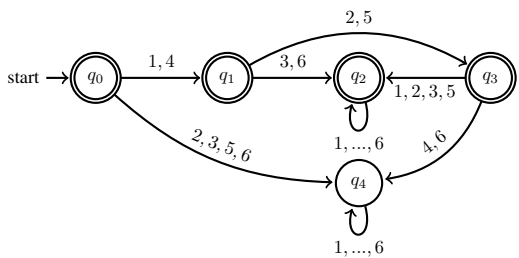

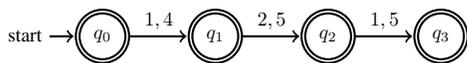
\begin{figure}[!t]
	\centering	
\begin{tikzpicture}[shorten >=1pt,node distance=2cm,on grid,auto, bend angle=20, thick,scale=0.75, every node/.style={transform shape}]
\node[state,initial,accepting] (q0)   {$q_0$};  
\node[state, accepting] (q1) [right=2cm of q0] {$q_1$}; 
\node[state, accepting] (q2) [right=2cm of q1] {$q_2$}; 
\node[state, accepting] (q3) [right=2cm of q2] {$q_3$}; 

\path[->]     
(q0) 
edge node[pos=0.5, sloped, above] {$1,4$} (q1) 

(q1) edge node [align=center] {$2,5$} (q2)	

(q2) edge node [align=center] {$1,5$} (q3)	
;	
\end{tikzpicture}
	\centering\caption{The witness za-DFA $\mathcal{F}_{c}$ in the fourth iteration}
	\label{fig::Fc4}
\end{figure}

In the fourth iteration, after adding $124$ to $Y$ and answering membership queries, we have the new observation table in Table \ref{tab::G4} and the za-DFA in Fig. \ref{fig::F4}. While both OracleP and OracleB return $true$, OracleB finds the adversary shown in Fig. \ref{fig::Fc4} that witnesses a probability of $0.3882>0.28$. This time OracleS returns $121$ as the negative counterexample which gives the satisfaction probability of $0.1179$. The updated $C_S=\{121,124\}$.

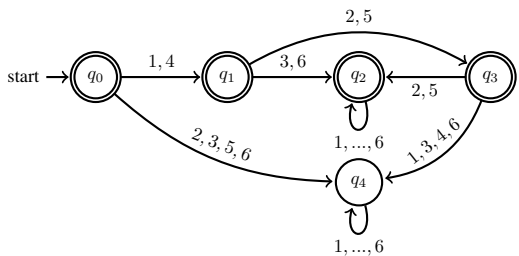
\begin{figure}[!t]
	\centering	
	\begin{tikzpicture}[shorten >=1pt,node distance=2cm,on grid,auto, bend angle=20, thick,scale=0.7, every node/.style={transform shape}]
	\node[state,initial,accepting] (q0)   {$q_0$};  
	\node[state, accepting] (q1) [right=2.5cm of q0] {$q_1$}; 
	\node[state, accepting] (q2) [right=2.5cm of q1] {$q_2$}; 
	\node[state, accepting] (q3) [right=2.5cm of q2] {$q_3$}; 
	\node[state] (q4) [below=2cm of q2] {$q_4$}; 
	\path[->]     
	(q0) 
	edge node[pos=0.5, sloped, above] {$1,4$} (q1) 
	edge [bend right = 20]  node[pos=0.5, sloped, above] {$2,3,5,6$} (q4) 
	(q1) edge node [align=center] {$3,6$} (q2)	
	edge[bend left = 30]  node[ pos=0.5, sloped, above] {$2,5$} (q3) 
	(q2) edge [loop below] node [align=center] {$1,...,6$} ()
	(q3) edge node [align=center] {$2,5$} (q2)	
	edge[bend left = 30]  node[ pos=0.5, sloped, above] {$1,3,4,6$} (q4)
	(q4) edge [loop below] node [align=center] {$1,...,6$} ()
	
	;
	\end{tikzpicture}
	\centering\caption{The acceptor za-DFA $\mathcal{F}_{5}$ in the sixth iteration}
	\label{fig::F5}
\end{figure}

In the next iteration, OracleS returns $123$ as the negative counterexample and $C_S=\{121,123,124\}$. After that, in the new iteration, we finally learn the za-DFA shown as Fig. \ref{fig::F5} with the maximum satisfaction probability $0.271<0.28$. Thus with six iterations, we have found the non-blocking and permissive supervisor that can regulate the closed-loop behavior to satisfy given PCTL specification.

\begin{table}[!t]
	\centering
	\caption{The observation table in the fourth iteration}
	\label{tab::G4}
	\begin{tabular}{c|cccc}
		G               & $\epsilon$ & 3 & 13 & 4 \\ \hline
		$\epsilon$      & 1          & 0 & 1  & 1 \\
		2               & 0          & 0 & 0  & 0 \\
		1               & 1          & 1 & 0  & 1 \\
		13              & 1          & 1 & 1  & 1 \\
		11              & 0          & 0 & 0  & 0 \\
		12              & 1          & 1 & 1  & 0 \\
		124             & 0          & 0 & 0  & 0 \\ \hline
		3               & 0          & 0 & 0  & 0 \\
		4               & 1          & 1 & 0  & 1 \\
		5               & 0          & 0 & 0  & 0 \\
		6               & 0          & 0 & 0  & 0 \\
		2$\{1,...,6\}$  & 0          & 0 & 0  & 0 \\
		14              & 0          & 0 & 0  & 0 \\
		15              & 1          & 1 & 1  & 0 \\
		16              & 1          & 1 & 1  & 1 \\
		13$\{1...,6\}$  & 1          & 1 & 1  & 1 \\
		11$\{1...,6\}$  & 0          & 0 & 0  & 0 \\
		12$\{1,2,3,5\}$ & 1          & 1 & 1  & 1 \\
		126             & 0          & 0 & 0  & 0 \\
		124$\{1...,6\}$ & 0          & 0 & 0  & 0
	\end{tabular}
\end{table}

\section{Conclusions and Future Work}

In this paper, an $L^*$ learning based supervisor synthesis framework is proposed for POMDP to satisfy formal specifications. With finite horizon PCTL being considered as system specifications, we design the supervisory control framework with za-DFA. By modifying the membership queries and conjectures in the $L^*$ algorithm, our learning process can automatically synthesize a za-DFA with termination guarantee. The proposed algorithm is also sound and complete. However, due to the challenges from the counterexample selection for probabilistic systems, OracleS may misidentify good control policies as counterexamples, in which cases the permissiveness of the supervisor cannot be guaranteed. 

In the future, we will explore different counterexample selection algorithms for probabilistic systems to reduce the possibility of misidentification from OracleS. While currently each part of the learning process is running separately based on different packages, for example, COMICS for DTMC counterexample selection \cite{jansen2012comics}, libalf for $L^*$ algorithm \cite{bollig2010libalf}, POMCP for POMDP solving \cite{silver2010monte}, we will glue every part together to deliver a whole software package for the automatic synthesis purpose.





\ifCLASSOPTIONcaptionsoff
  \newpage
\fi



\bibliographystyle{IEEEtran}
\bibliography{bib/ref,bib/ref1,bib/ref3,bib/ref2,bib/Cyber-RelatedWork1,bib/extra} 
%

%







\end{document}